\documentclass[]{spie}  

\usepackage[dvips]{graphicx}

\def\degC{{${}^\circ$C}}
\def\degree{{${}^\circ$}}

\begin{document}

\thispagestyle{empty}
\renewcommand{\thefootnote}{\fnsymbol{footnote}}

\begin{flushright}
{\small
SLAC--PUB--9493\\
ISAS--RN--754\\
September, 2002\\}
\end{flushright}

\vspace{.8cm}

\begin{center}
{\bf\large   
Low Noise Double-Sided Silicon Strip Detector \\
for Multiple-Compton Gamma-ray Telescope
\footnote{Work supported by
Department of Energy contract  DE--AC03--76SF00515, Grantin-Aid by Ministry of Education, Culture, Sports, Science and Technology of Japan (12554006, 13304014),
and ``Ground-based Research Announcement for Space Utilization'' promoted by Japan Space Forum.}}

\vspace{1cm}

Hiroyasu Tajima and Tuneyoshi Kamae\\
Stanford Linear Accelerator Center, Stanford University,
Stanford, CA  94309\\

\medskip

Shingo Uno, Tatsuya Nakamoto and Yasushi Fukazawa\\
Department of Physics, Hiroshima University, 
Higashi-Hiroshima 739-8526, Japan\\

\medskip

Takefumi Mitani, Tadayuki Takahashi and Kazuhiro Nakazawa\\
Institute of Space and Astronautical Science, 
Sagamihara, Kanagawa 229-8510, Japan\\

\medskip

Yu Okada\\
Department of Physics, University of Tokyo, 
Bunkyo-ku, Tokyo 113-0033, Japan\\

\medskip

Masaharu Nomachi\\
Department of Physics, Osaka University, Toyonaka, Osaka 560-0043, Japan\\

\end{center}

\vfill

\begin{center}
{\bf\large   
Abstract }
\end{center}

\begin{quote}
A Semiconductor Multiple-Compton Telescope (SMCT) is being developed to explore the gamma-ray universe in an energy band 0.1--20~MeV, which is not well covered by the present or near-future gamma-ray telescopes. 
The key feature of the SMCT is the high energy resolution that is crucial for high angular resolution and high background rejection capability.
We have developed prototype modules for a low noise Double-sided Silicon Strip Detector (DSSD) system which is an essential element of the SMCT.
The geometry of the DSSD is optimized to achieve the lowest noise possible. 
 A new front-end VLSI device optimized for low noise operation is also developed.
We report on the design and test results of the prototype system. 
We have reached an energy resolution of 1.3~keV (FWHM) for 60~keV and 122~keV at 0\degC.
\end{quote}

\vfill

\begin{center} 
{\it Contributed to} 
{\it Astronomical Telescopes and Instrumentation }\\
{\it Waikoloa, Hawaii USA}\\
{\it August 22--August 28, 2002} \\



\end{center}

\newpage



%
\pagestyle{plain}

\section{Introduction}
\label{sect:intro}  

The gamma-ray universe in the energy band above 0.1~MeV provides a rich ground to study nucleosynthesis and physics of particle acceleration beyond thermal emission.
However, the energy band between 0.1~MeV and 100~MeV is poorly explored due to difficulties associated with the detection of such photons.
The Compton telescope COMPTEL\cite{COMPTEL93} abroad CGRO (Compton Gamma-Ray Observatory) demonstrated that a gamma-ray instrument based on the Compton scattering is useful for the detection of the gamma-ray in this energy band.
COMPTEL provided us with rich information on a variety of gamma-ray emitting objects either in continuum and line emission. 
The continuum sources include spin-down pulsars, stellar black-hole candidates, supernovae remnants, interstellar clouds, active galactic nuclei (AGN), gamma-ray bursts (GRB) and solar flares. 
Detection has also been made of the nuclear gamma-ray lines from ${}^{26}$Al (1.809~MeV), ${}^{44}$Ti (1.157~MeV), and ${}^{56}$Co (0.847 and 1.238~MeV). 

Although COMPTEL performed very well as the first Compton telescope in space for MeV gamma-ray astrophysics, it suffered severely from large background, poor angular resolution, and complicated image decoding.\cite{Knodlseder96}
In 1987, T. Kamae {\it et al}. proposed a new Compton telescope based on a stack of silicon strip detectors (SSD).\cite{Kamae87,Kamae88}
This technology presents very attractive possibilities to overcome the weaknesses of COMPTEL as described later in this document.
This idea of using silicon strip detectors stimulated new proposals for the next generation Compton telescope.\cite{Takahashi,MEGA,Milne}

Recently, a new semiconductor detector based on CdTe (Cadmium Telluride) emerged as a promising detector technology for detection of MeV gamma-rays.\cite{Takahashi01}
Taking advantage of significant development in CdTe technology, we have proposed a new generation of Compton telescope, a Semiconductor Multiple-Compton Telescope (SMCT).
The SMCT is a hybrid semiconductor gamma-ray detector which consists of  silicon and CdTe detectors.
Excellent energy resolution is the key feature of the SMCT to achieve high angular resolution and background rejection capability.
In addition,  the ability to measure polarization, a wide energy band (0.1--20 MeV), a wide field of view ($\sim$60\degree) and light weight are important features of the SMCT.
Low noise Double-sided Silicon Strip Detectors (DSSDs) are fundamental elements of the SMCT.
In this paper, we describe the optimization of the detector components to realize the fine energy resolution and the test results with prototype modules.
Other applications for such low noise DSSD system and hybrid Si-CdTe system are described elsewhere.\cite{Takahashi02}


\section{Multiple-Compton technique}

A simple Compton telescope consists of two components, a scatterer and an absorber. 
The direction of the gamma-ray can be confined in a cone by measuring the energy loss and position of the scattering in the scatterer and those of photoelectric-absorption in the absorber.
In order to increase the detection efficiency of the gamma-ray, the scatterer has to be thicker and/or more dense.
However, it also increases  the probability of multiple-Compton scatterings in the scatterer, which was considered to be impossible to resolve.
In order to solve this problem, a stack of many thin scatterers is used to record individual Compton scatterings in the Multiple-Compton technique.
The order of the interaction sequences, hence the correct energy and direction of the incident photon, can be reconstructed by examination of the energy-momentum conservation for all possible sequences. 
This technique has proven to be a very powerful method to suppress backgrounds. 

   \begin{figure}[tbh]
   \begin{center}
   \begin{tabular}{ll}
   (a) & (b) \\
          & \includegraphics[height=4.5cm]{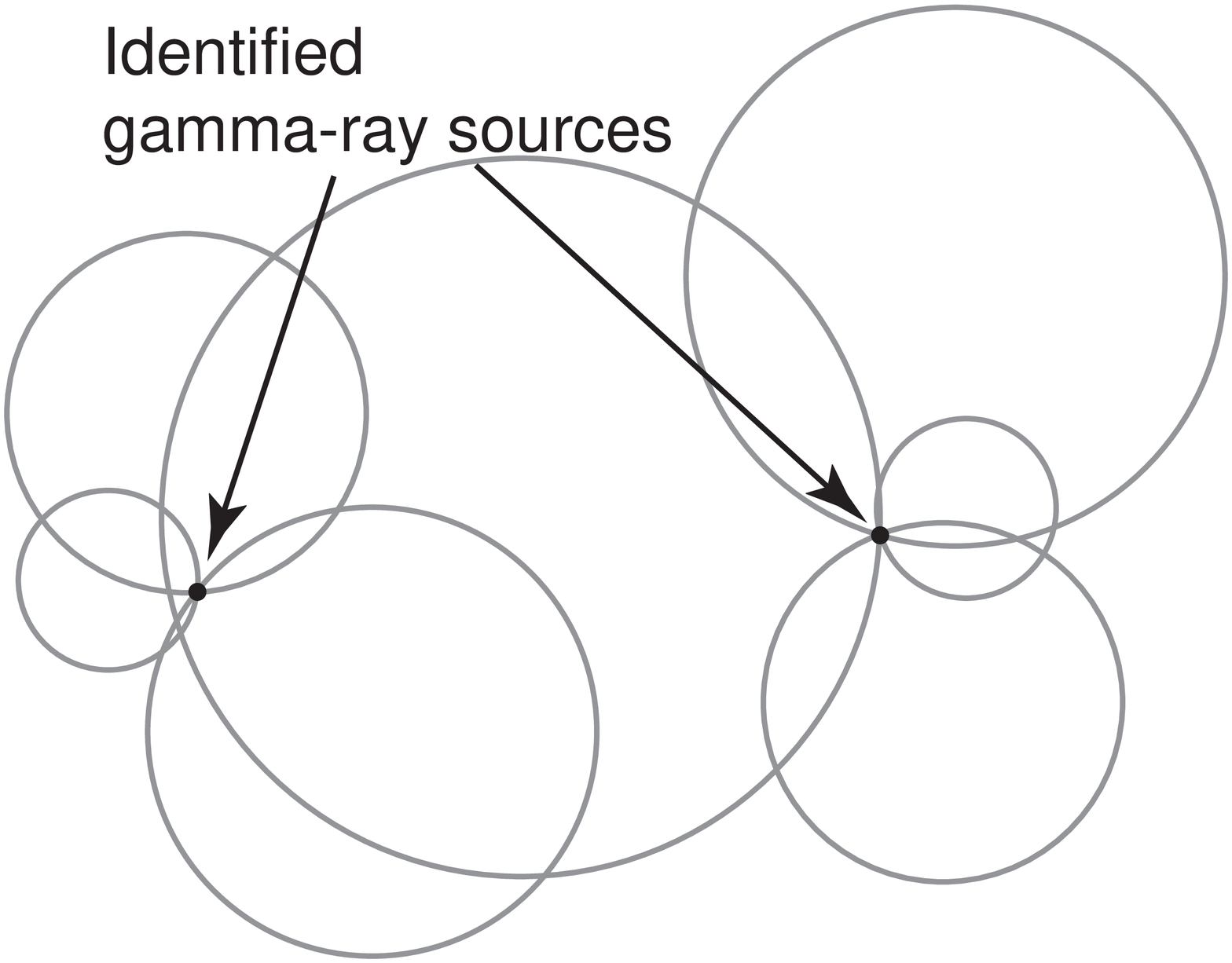} \\
          & (c) \\
\raisebox{1.0cm}[0pt]{\includegraphics[height=8cm]{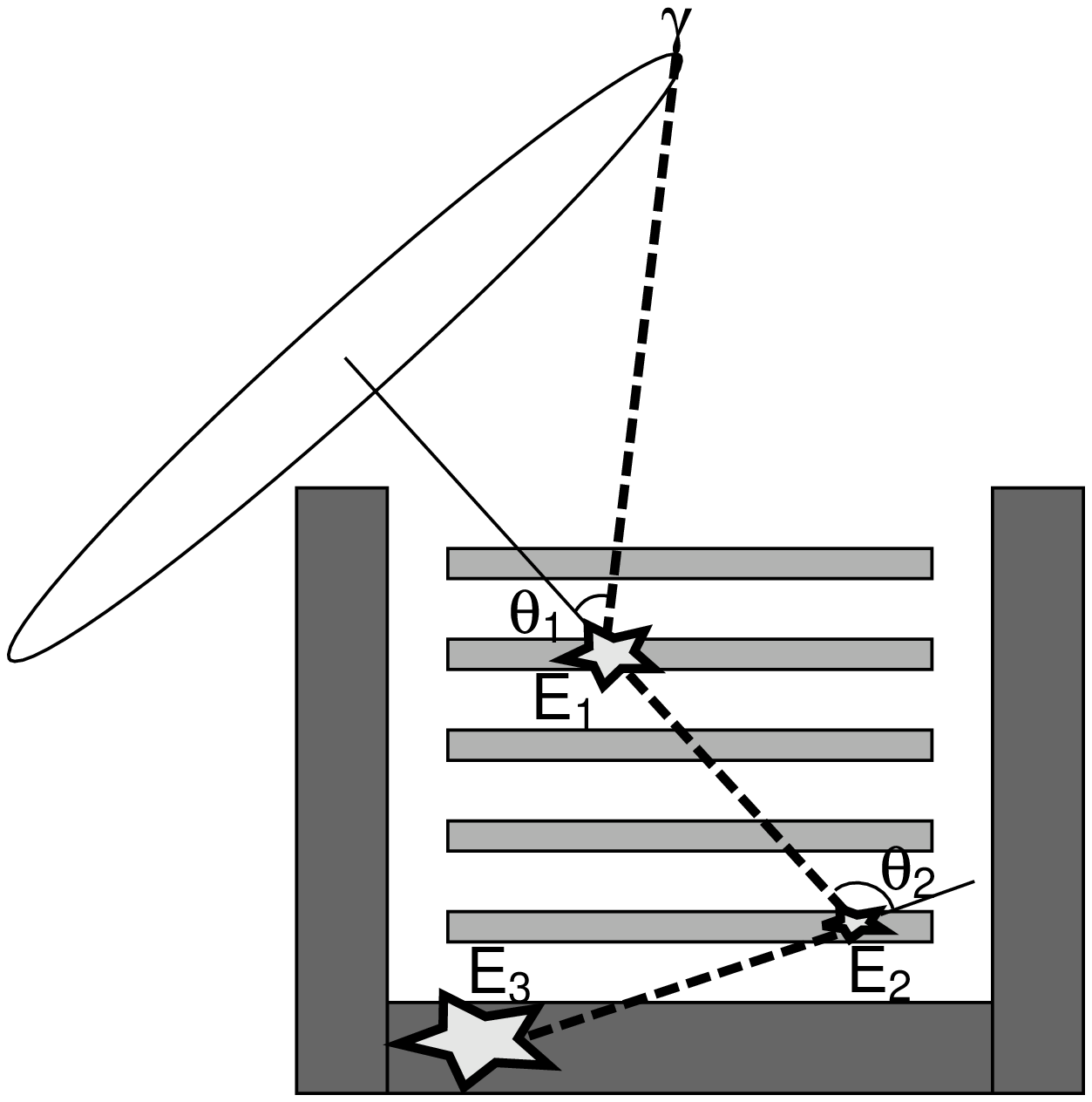} \hspace*{1.5cm}}
          & \includegraphics[height=4.5cm]{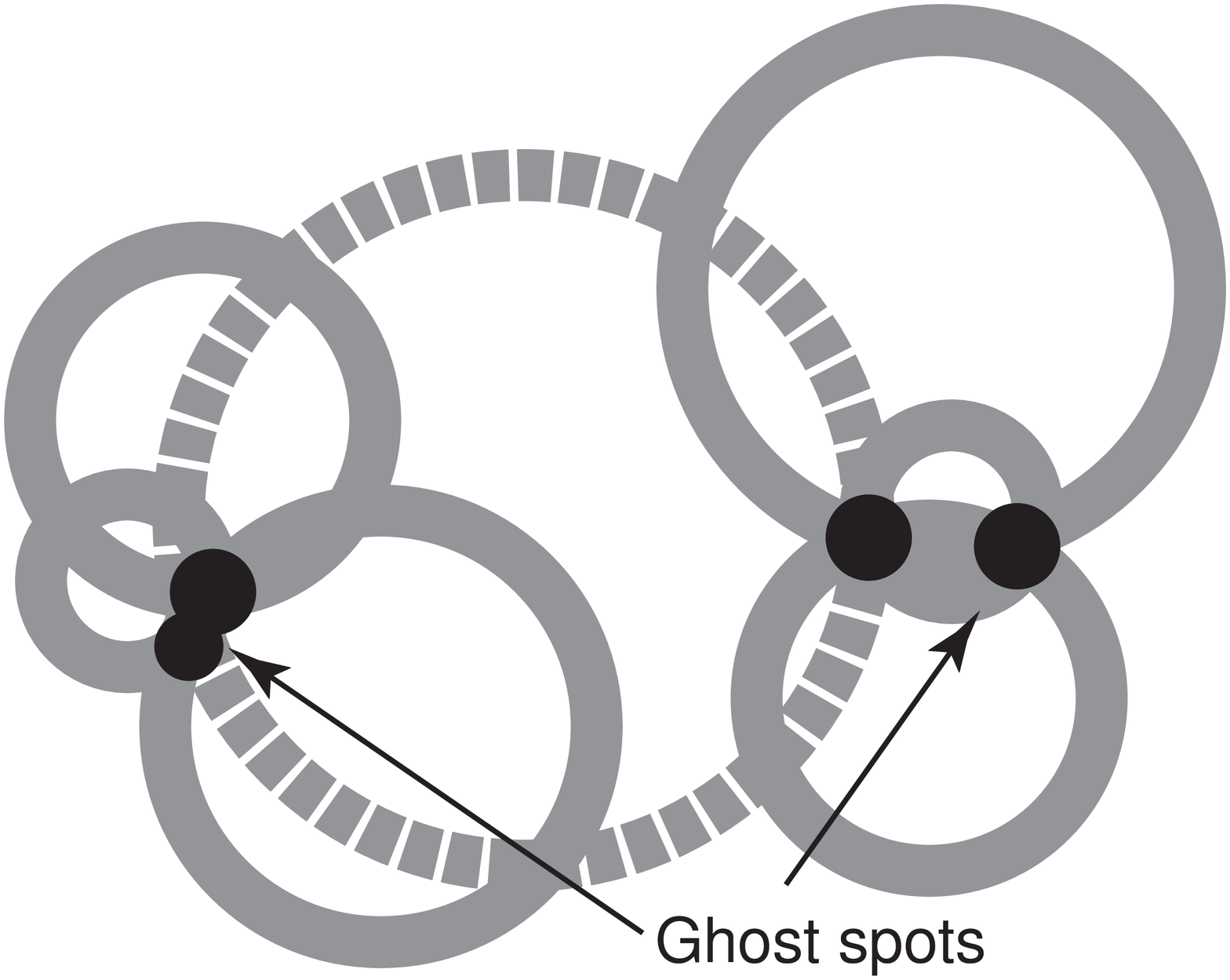}\\
   \end{tabular}
   \end{center}
   \caption[Concept of Multiple-Compton technique.] 
   { \label{fig:schematic} (a) Concept of Multiple-Compton technique. (b) and (c) Illustration of processes to reconstruct source locations using photon candidates from two gamma-ray sources. The angular resolution in (b) is 10 times better than that in (c).}
   \end{figure} 
Figure~\ref{fig:schematic} (a)  illustrates a case with two Compton scatterings in the scatterer and one photoelectric-absorption in the absorber. 
In this situation, both scattering angles $\theta_1$  and  $\theta_2$  can be obtained from the recoil electron energies as
\begin{equation}
\cos\theta_1 = 1+\frac{m_ec^2}{E_1+E_2+E_3}-\frac{m_ec^2}{E_2+E_3},\ \ 
\cos\theta_2 = 1+\frac{m_ec^2}{E_2+E_3}-\frac{m_ec^2}{E_3},
\end{equation}
where $E_1$, $E_2$ and $E_3$ are the energy deposits in each photon interaction.
It is worthwhile noticing that $\theta_2$ can be reconstructed from the hit positions of the three interactions. 
This over-constraint provides stringent background suppression that is crucial to observe faint or diffuse gamma-ray sources. 
Note that the photon does not have to be absorbed completely in this multiple-Compton technique since the incident photon energy can be inferred when three or more Compton scatterings are recorded, which means no need for a heavy calorimeter in the telescope. 
Another possible advantage of this technique is that recoil electrons with energies more than 250~keV penetrate a 300-$\mu$m silicon detector and deposit energies in two layers of silicon detectors. 
Although the angular resolution is very poor for such low energy electrons due to multiple-Coulomb scattering, this information can still be used to constrain the direction of the incident photon to a partial cone.

High energy resolution for the recoil electron is the critical characteristic of Compton telescope since it determines the angular resolution of the incident photon, and background rejection capability using Compton kinematics.
Figures~\ref{fig:schematic} (b) and (c) illustrate processes to reconstruct gamma-ray source locations with different angular resolutions to demonstrate the importance of angular resolution.
In these figures, a circle represents a projection of a cone reconstructed by the Compton kinematics onto a sky map for each photon candidate.\footnote{The projection is not actually  a circle. We use a circle for simplicity.}
The width of the circles in figure~\ref{fig:schematic} (b) is 10 times narrower than that in figure~\ref{fig:schematic} .
We can clearly identify two gamma-ray sources in figure~\ref{fig:schematic} (b).
In figure~\ref{fig:schematic} (c), it is difficult to resolve them because ghost spots cannot be rejected due to poor resolution.
In addition, the dashed circle can belong to two spots from different sources, further complicating the matter.
Since the circle radius can be very large depending on the Compton scattering angle, we may suffer large background from any bright source in the field of view.
Therefore,  precise angular resolution is essential to resolve source locations and to suppress background from bright sources.

   \begin{figure}[htb]
   \begin{center}
   \begin{tabular}{c}
   \includegraphics[height=7.2cm]{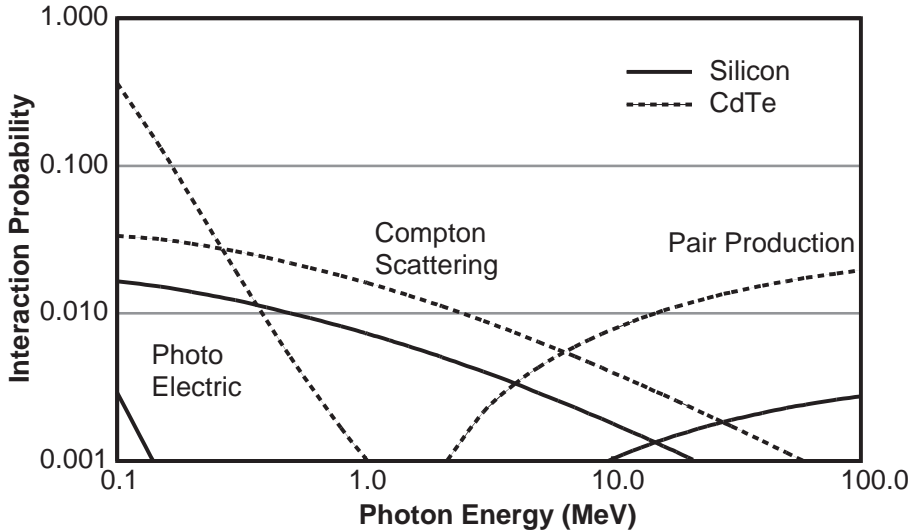}
   \end{tabular}
   \end{center}
   \caption[Concept of Multiple-Compton technique.] 
   { \label{fig:probability} Probability of various photon interactions in 500 $\mu$m thick silicon and CdTe devices as a function of the photon energy.}
   \end{figure} 
The DSSD is a primary candidate for the scatterer because Compton scattering dominates over other processes in sub-MeV range and the Doppler-broadening effect is small in silicon.
Such double-sided devices are required to provide two dimensional measurement of the interaction position.
Weak stopping power of the silicon detector at the upper end of the energy band can be supplemented by employing the CdTe detectors in the back portion of the layers.
Figure~\ref{fig:probability} shows the probability of various photon interactions in 500 $\mu$m thick silicon and CdTe devices as a function of the photon energy. 
This figure clearly illustrates the advantage of a Si-CdTe hybrid approach to detect photons in a wide energy-band.
For the photon energies below 0.5~MeV, silicon is a better scatterer since the Compton scattering cross section is larger than that of the photoelectric absorption.
Note that photoelectric absorption does not provide information on the incident photon direction.
The DSSD layers measure the recoil electron energy of the first Compton scattering with maximal energy resolution while the CdTe layers enhance the probability of Compton scattering and/or photoelectric absorption in the back section. 
At higher energies, silicon becomes more or less transparent and CdTe detector becomes the dominant detector.
Pair creation can also be utilized to measure both the energy and direction of gamma-rays above several MeV.
However, the efficiency is significantly reduced above 20 MeV since the gamma-ray cannot be completely absorbed.
In this case, a higher Z material is required to enhance photon absorption.

As discussed above, precise energy resolution is a crucial feature of the DSSDs to gain high pointing accuracy and large background rejection capability, utilizing the over-constraint from the kinematics of multiple-Compton scattering.
Low electronics noise is required to accomplish such fine energy resolution by exploiting the small Doppler-broadening effect in silicon.
A target noise value in equivalent noise charge (ENC) for the DSSD system is 100~$e^-$ (RMS, Root Mean Squared), which corresponds to 1~keV (FWHM, Full-Width Half Maximum) for the silicon detector.
This equates to an angular resolution of 1\degree\ at 30~keV or 0.3\degree\ at 2~MeV.
Angular resolution at the upper end of the energy band is dominated by the position resolution of the detector.

Possible instrument configurations and expected performance are discussed in ref~\citenum{Takahashi02}.

\section{Low Noise SSD System}

We have developed prototype modules for a low noise DSSD system in order to understand all noise sources in detail, which is fundamental to achieve the best possible energy resolution.
A low noise DSSD system consists of a DSSD, an RC chip and a VA32TA front-end VLSI chip.
The DSSD does not employ an integrated AC capacitor in order to maintain a strip yield close to 100\%.
The RC chip provides detector bias voltage via polysilicon bias resistors, as well as AC-coupling between strips and preamplifier channel inputs.
Since the RC chip is a small device, the yield of the AC-coupling capacitor is not a concern.
The VA32TA front-end VLSI is a 32-channel, low noise CMOS device which includes on a per-channel basis a preamplifier, shaper, sample/hold, analog multiplexer and discriminator.
The DSSDs and RC chips are manufactured by Hamamatsu Photonics, Japan and the VA32TA is manufactured by Ideas ASA, Norway.
Detailed description of each component is given below.

   \begin{figure}[tbh]
   \begin{center}
   \begin{tabular}{ll}
(a) DSSD junction side & (b) DSSD ohmic side \\
 \includegraphics[height=5.5cm]{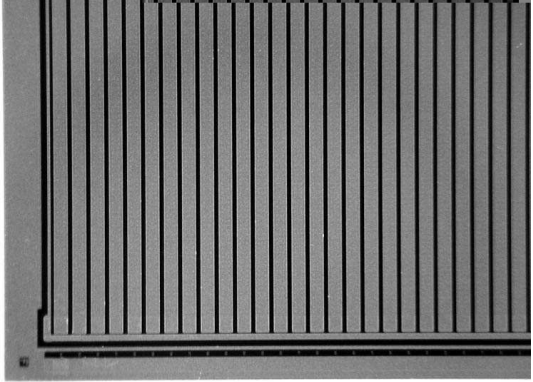} \hspace*{1cm} &
 \includegraphics[height=5.5cm]{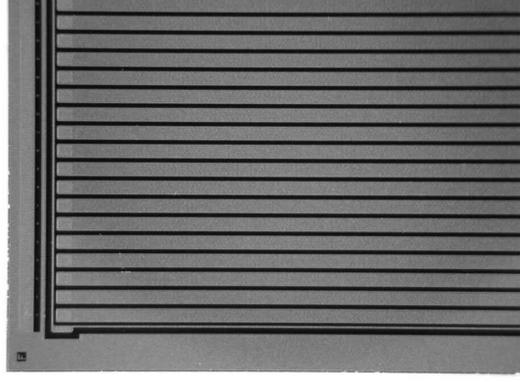}
   \end{tabular}
   \end{center}
   \caption[Pictures of a DSSD for (a) junction side and (b) ohmic side. Strips are surrounded by guard ring.] 
   { \label{fig:DSSD} Pictures of a DSSD for (a) junction side and (b) ohmic side. Strips are surrounded by guard ring.}
   \end{figure} 
The p${}^+$-strips are implanted on an n${}^-$ silicon substrate to form p-n junctions. 
This side is called junction side.
The n${}^+$-strips are implanted orthogonal to the p-strips on the other side to provide second coordinate measurement. This side is called ohmic side.
The p${}^+$-implant surrounds the individual n${}^+$-strip to provide separation of n${}^+$-strips.
Figure~\ref{fig:DSSD} shows pictures of the junction and ohmic sides of the DSSD.
Because the strip geometry is more complex on the ohmic side than on the junction side, the strip capacitance is larger on the ohmic side.
For this reason, the primary function of the ohmic side is to provide position information of the second coordinate, and its noise performance is not measured in this test.
We have produced five types of strip geometries in four types of devices (400-1, 400-2, 800-1, 800-2) as listed in Tables~\ref{tab:DSSD} and \ref{tab:SSD} to understand the noise dependence on parameters such strip pitch, strip gap and strip length.
Wider strip pitch gives larger capacitance to the backplane and larger leakage current.
Wider strip gap gives smaller capacitance to the adjacent strips although the risk of junction breakdown increases at high bias voltage.
The 400-2a $\sim$ 400-2c configurations are implemented on the junction side of a single-sided SSD. (Ohmic side does not have strip structure.)  
All four types of devices are implemented on one silicon wafer.
The thickness of the silicon wafer is fixed at 300~$\mu$m.
Here we present results on the basic characteristics for  types 400-2a, 400-2b and 400-2c SSDs.
The C-V curve measurement yields the depletion voltage of 65~V, therefore the following measurements are performed at 70~V.
\begin{table}[bth]
\caption{Parameters for strip geometry of the double-sided SSD prototypes.} 
\label{tab:DSSD}
\begin{center}       
\begin{tabular}{|r|r|r|r|} 
\hline
\rule[-1ex]{0pt}{3.5ex}  Type & Strip pitch [$\mu$m] & Strip gap [$\mu$m]  & Strip length [cm]  \\
\hline\hline
\rule[-1ex]{0pt}{3.5ex} 400-1 & 400 &  100 & 2.56  \\
\hline
\rule[-1ex]{0pt}{3.5ex}  800-1 & 800 &  100 & 2.56  \\
\hline
\rule[-1ex]{0pt}{3.5ex}  800-2 & 800 &  100 & 1.27  \\
\hline
\end{tabular}
\end{center}
\end{table}
\begin{table}[bth]
\caption{Parameters for strip geometry of the single-sided SSD prototype.} 
\label{tab:SSD}
\begin{center}       
\begin{tabular}{|r|r|r|r|} 
\hline
\rule[-1ex]{0pt}{3.5ex}  Type & Strip pitch [$\mu$m] & Strip gap [$\mu$m]  & Strip length [cm]  \\
\hline\hline
\rule[-1ex]{0pt}{3.5ex} 400-2a & 400 &  100 & 2.56  \\
\hline
\rule[-1ex]{0pt}{3.5ex}  400-2b & 400 &  130 & 2.56  \\
\hline
\rule[-1ex]{0pt}{3.5ex}  400-2c & 400 &  160 & 2.56  \\
\hline
\end{tabular}
\end{center}
\end{table}
A bias voltage of up to 200~V was applied to test junction breakdown and no breakdown was observed. 
Leakage current is measured to be 0.5~nA/strip at 20\degC\ and 0.05~nA/strip at 0\degC.
The strip capacitance is measured to be $6.3\pm0.2$~pF for 400-2a, $5.7\pm0.2$~pF for 400-2b and $5.2\pm0.2$~pF for 400-2c, which is somewhat larger than the values calculated based upon geometry of 5.1~pF, 4.8~pF and 4.5~pF, respectively.
In the capacitance calculations, we take into account only capacitance to the backplane and adjacent strips.
Measured values of the strip capacitance are used for the subsequent noise analysis.

The RC chip consists of bias resistors and AC-coupling capacitors.
We employ 1 G$\Omega$ polysilicon resistor for the bias resistor.
A resistance value of 1 G$\Omega$ is chosen to minimize thermal noise without compromising production stability.
A MOS capacitor is employed for AC-coupling to ensure high voltage tolerance, since it needs to withstand the bias voltage of the DSSD and to protect the preamplifiers.
The size of this MOS capacitor is 1.2~mm $\times$ 0.14~mm in order to obtain sufficient coupling capacitance while presenting a capacitance load to the preamplifier of less than 1~pF.

   \begin{figure}[tbh]
   \begin{center}
   \begin{tabular}{c}
   \includegraphics[height=4cm]{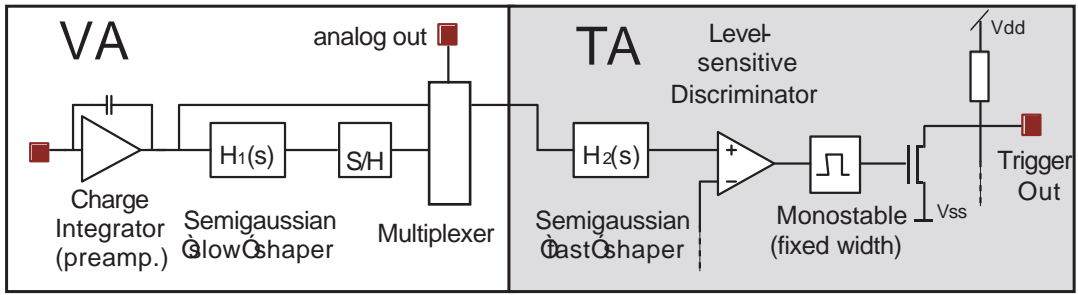}
   \end{tabular}
   \end{center}
   \caption[Block diagram of VA32TA preamplifier LSI.] 
   { \label{fig:VA32TA} Block diagram of VA32TA front-end LSI.}
   \end{figure} 
We have developed the VA32TA front-end ASIC based on the VA32C amplifier VLSI and the TA32C trigger VLSI that are originally developed by Ideas.
A detailed description of Viking-architecture chip is given elsewhere.\cite{VA91,VA94}
The VA32TA is  fabricated in the AMS 0.35 $\mu$m technology, which is measured to be radiation tolerant to 20~MRad or more.\cite{Yokoyama01}
A VA32TA consists of 32 channels of signal-readout.  
Each channel includes a charge sensitive preamplifier-slow CR-RC shaper-sample/hold-analog multiplexer chain (VA section) and fast shaper-discriminator chain (TA section) as illustrated in the block diagram of Figure~\ref{fig:VA32TA}.
The front-end MOSFET size for the preamplifier is optimized for small capacitance load, to attain optimal noise performance.
Expected noise performance is $(45+19\times C_d)/\sqrt{\tau}\;e^-$ (RMS) in ENC or $(0.37+0.16\times C_d)/\sqrt{\tau}$~keV (FWHM), where $C_d$ is the load capacitance and $\tau$ is the peaking time, which can be varied from 1 to 4 $\mu$s.
Feedback resistors for the preamplifier, as well as slow and fast shapers are realized with MOSFETs.
Gate voltages of the feedback MOSFETs are controlled by internal DACs (Digital-to-Analog Converters) on chip.
Bias currents for various components are also controlled by the internal DACs.
Threshold levels can be adjusted for each channel using individual DACs to minimize threshold dispersion.
A 200-bit register is required to hold the values for all internal DACs.
Majority selector logic circuitry has been utilized for these registers to ensure the tolerance against Single-Event Upset (SEU), which is important for space applications.
This majority selector circuitry uses three flip-flops for each bit and takes a majority of the three when they become inconsistent.
This logic also generates a signal when such inconsistencies are detected.

\begin{figure}[tbh]
   \begin{center}
   \begin{tabular}{ll}
(a) AC configuration & (b) DC configuration \\
   \includegraphics[height=3.5cm]{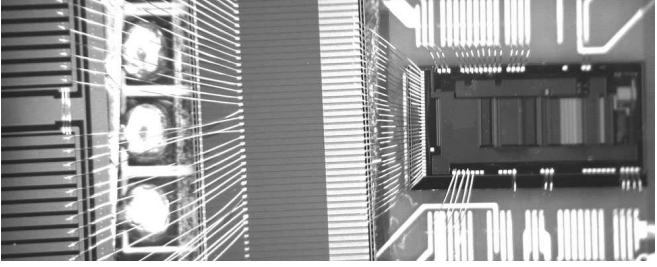} \hspace*{0.5cm} &
   \includegraphics[height=3.5cm]{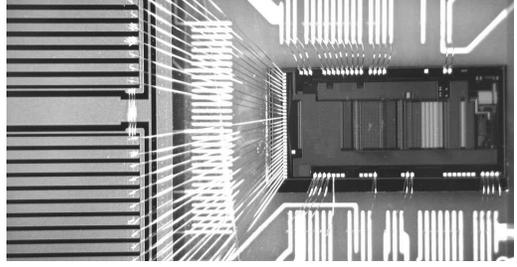}
   \end{tabular}
   \end{center}
\caption[Pictures of prototype SSD modules with (a) AC and (b) DC configurations. 
The SSD is located on the left and the VA32TA is located on the right.
In the AC configuration, an RC chip is present between the SSD and VA32TA.] 
   { \label{fig:module} Pictures of prototype SSD modules with (a) AC and (b) DC configurations.  
The DSSD is located on the left and the VA32TA is located on the right.
In the AC configuration, an RC chip is present between the SSD and VA32TA.}
   \end{figure}

We have assembled two prototype modules, one consists of SSD, RC chip and VA32TA (AC module) and the other module consists of SSD and VA32TA (DC module).
Type 400-2 SSD is used for both modules.
Figures~\ref{fig:module} (a) and (b) show pictures of the AC and DC modules, respectively.
Only 400-2b and 400-2c strips are connected to the readout system because of geometric constraints.
Only the junction side is connected to the readout system in this test since energy resolution is determined by the noise performance on the junction side.
The ohmic side is known to give worse noise performance due to larger strip capacitance and used primarily to provide the coordinate information and a trigger coincidence.

\section{ Noise Performance }
We have taken into account the following noise sources in our analysis to optimize the component parameters:
\begin{itemize}
\item Preamplifier noise: linear dependence on the capacitance load and $1/\sqrt{\tau}$ dependence on the peaking time. 
Noise level varies from 0.8~keV (DC, $\tau=4$ $\mu$s) to 1.4~keV (AC, $\tau=2$ $\mu$s).
\item Shot noise: $\sqrt{I}$ dependence on the leak current and $\sqrt{\tau}$ dependence on the peaking time.
Noise level varies from 0.3~keV (0\degC, $\tau=2$ $\mu$s) to 1.2~keV (20\degC, $\tau=4$ $\mu$s).
\item Thermal noise (bias resistor): $1/\sqrt{R}$ dependence on the bias resistance and $\sqrt{\tau}$ dependence on the peaking time.
Noise level varies from 0.3~keV ($\tau=2$ $\mu$s) to 0.4~keV ($\tau=4$ $\mu$s).
\end{itemize}
We have also considered the thermal noise from the implant resistance of the RC chip and found it is less than 0.1~keV, hence considered negligible.
Total noise is fairly independent of the peaking time beyond 2~$\mu$s because amplifier noise and other noise sources show opposite peaking time dependence and compensate each other.

Noise performance of the prototype system is measured at temperatures of 0\degC\ and 20\degC\ and at peaking times of 2~$\mu$s and 4~$\mu$s.
This is useful to differentiate the noise contributions.
Absolute gain of the system is calibrated using the X-ray spectra obtained below.
Table~\ref{tab:resolution} summarizes the measurement results and compares them with calculation results.
This table does not include the result for the DC configuration at 20\degC\ since it cannot be operated at optimum condition due to the effect of leak current from the SSD flowing into the preamplifier.
Since the results for 400-2b and 400-2c do not show significant difference (less than 5\% though the 400-2c consistently gives better noise performance), the average of the two types are shown in this paper.
\begin{table}[tbh]
\caption{Noise and energy resolution measured for  the prototype modules in the AC and DC configurations at temperatures of 0\degC\ and 20\degC\ and peaking times of 2~$\mu$s and 4~$\mu$s.} 
\label{tab:resolution}
\begin{center}       
\begin{tabular}{|c|r|r||r|r||r|r|} 
\hline
\rule[-1ex]{0pt}{3.5ex}   &  & Peaking & \multicolumn{2}{c||}{ Noise (FWHM)} &  \multicolumn{2}{c|}{Energy resolution (FWHM)} \\
\cline{4-7}
\rule[-1ex]{0pt}{3.5ex} \raisebox{1.5ex}[0pt]{Configuration} & \raisebox{1.5ex}[0pt]{Temperature}  &  \multicolumn{1}{c||}{time} & \multicolumn{1}{c|}{Expected} &  \multicolumn{1}{c||}{Measured} & \multicolumn{1}{c|}{60~keV (${}^{241}$Am)} & \multicolumn{1}{c|}{122~keV (${}^{57}$Co)}\\
\hline\hline
\rule[-1ex]{0pt}{3.5ex} &  & 2 $\mu$s  & 1.6~keV & 1.7~keV & 1.9~keV & 2.1~keV\\
\rule[-1ex]{0pt}{3.5ex}  & \raisebox{1.5ex}[0pt]{20\degC} & 4 $\mu$s  & 1.6~keV & 1.7~keV & 2.1~keV & 2.3~keV\\
\cline{2-7}
\rule[-1ex]{0pt}{3.5ex}  \raisebox{1.5ex}[0pt]{AC} &  & 2 $\mu$s  & 1.4~keV & 1.6~keV & \hspace*{0.8cm}1.8~keV  & \hspace*{0.8cm}1.9~keV\\
\rule[-1ex]{0pt}{3.5ex} & \raisebox{1.5ex}[0pt]{0\degC} & 4 $\mu$s  & 1.1~keV & 1.4~keV & 1.7~keV & 1.8~keV\\
\hline
\rule[-1ex]{0pt}{3.5ex} &  & 2 $\mu$s  & 1.1~keV & 1.2~keV & 1.6~keV & 1.6~keV\\
\rule[-1ex]{0pt}{3.5ex} \raisebox{1.5ex}[0pt]{DC} & \raisebox{1.5ex}[0pt]{0\degC} & 4 $\mu$s  & 0.9~keV & 1.0~keV & 1.3~keV & 1.3~keV\\
\hline
\end{tabular}
\end{center}
\end{table}
These results confirm that the shot noise due to the leakage current becomes negligible at 0\degC.
In the DC configuration, the measured and expected noise values are in a good agreement for the measurements at either peaking time, which demonstrates that the noise sources from the SSD and the VA32TA are well understood.
On the other hand, slight disagreement between measurement and calculation is observed for the AC configuration at 0\degC, indicating additional noise sources in the RC chip which are not accounted for in the noise analysis.
At 20\degC\ the shot noise from the leak current becomes large and the disagreement becomes less significant.
Further studies are required to identify the origin of the excess noise observed in the RC chip in this measurement.

\section{ X-ray energy resolution}
The energy resolution for the X-ray detection is investigated using the 59.54~keV X-ray line from ${}^{241}$Am and the 122.06~keV X-ray line from ${}^{57}$Co.
Absolute gain is calibrated for each channel using the same X-ray sources in such way that all channels give the correct peak height.
We also correct for the gain dependence on the``common mode shift" where common mode shift refer to the voltage shift common to all channels in one chip.\footnote{The coomon mode shift is calculated and subtracted for each event taking the average of the pulse height for every channel, excluding the channels with signal.}
We attribute the origin of this dependence to the nonlinearity of the amplifier gain.
A correction of approximately 0.2~keV is made for a 1~keV common mode shift.
This correction is critical to operate the system at 20\degC, since the magnitude of the common mode noise is $6\sim10$~keV at 20\degC\ and 1.6~keV at 0\degC.
Larger common mode noise is observed for the longer peaking time at 20\degC, implying a low frequency nature of the common mode noise.
It should be noted that the peaking time of the fast shaper in the TA section is 0.3~$\mu$s, which helps to suppress the contribution from common mode noise.

\begin{figure}[tbh]
   \begin{center}
   \begin{tabular}{ll}
(a) ${}^{241}$Am spectrum & (b) Fit on the 59.5~keV X-ray spectrum.\\
   \includegraphics[height=5.4cm]{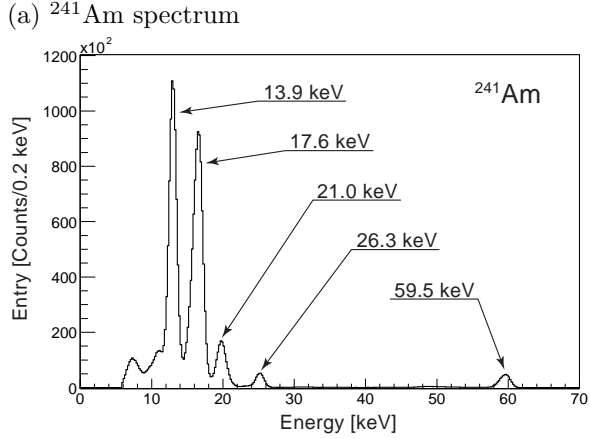} \hspace*{0.2cm} &
   \includegraphics[height=5.2cm]{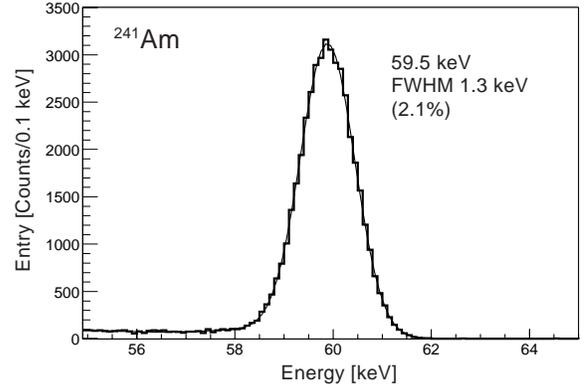}
   \end{tabular}
   \end{center}
   \caption[(a) Energy spectrum of ${}^{241}$Am measured by the DC module at 0\degC. (b) Magnified view around the 59.5~keV X-ray peak. The curve represents the fit result described in the text.] 
   { \label{fig:Am} (a) Energy spectrum of ${}^{241}$Am measured by the DC module at 0\degC. (b) Magnified view around the 59.5~keV X-ray peak. The curve represents the fit result described in the text.}
   \end{figure}

\begin{figure}[tbh]
   \begin{center}
   \begin{tabular}{ll}
(a) ${}^{57}$Co spectrum & (b) Fit on the 122.1~keV X-ray spectrum. \\
   \includegraphics[height=5.2cm]{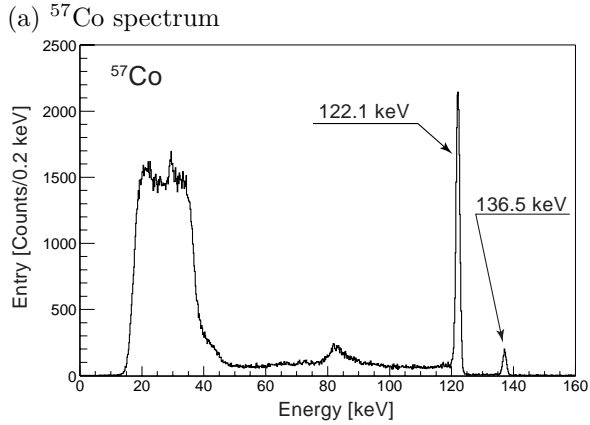} \hspace*{0.8cm} &
   \includegraphics[height=5.2cm]{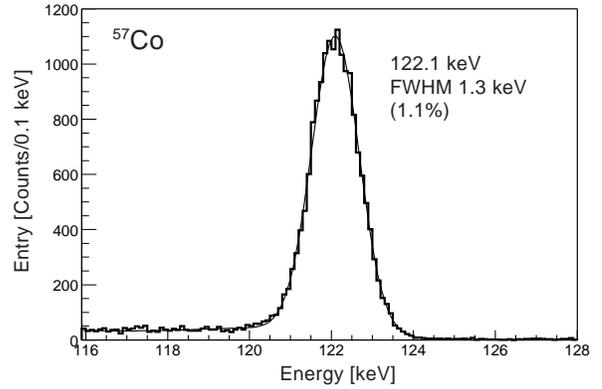}
   \end{tabular}
   \end{center}
   \caption[(a) Energy spectrum of ${}^{57}$Co measured by the DC module at 0\degC. (b) Magnified view around the 122.1~keV X-ray peak. The curve represents the fit result described in the text.] 
   { \label{fig:Co} (a) Energy spectrum of ${}^{57}$Co measured by the DC module at 0\degC. (b) Magnified view around the 122.1~keV X-ray peak. The curve represents the fit result described in the text.}
   \end{figure}

Figures~\ref{fig:Am} (a) and \ref{fig:Co} (a) show the sum of the  energy spectra for all channels, except for the first and last strips where we observe larger noise.
We observe a clear Compton edge just below 40~keV in the ${}^{57}$Co spectrum.
Peak positions other than 59.5~keV and 122.1~keV are somewhat shifted since gain calibration has not been performed for these energy points yet.
In order to obtain the energy resolution, we fit the energy spectrum to a sum of two Gaussian functions to describe the peak and one threshold type function\footnote{This function is expressed as $f(E) = p_1\cdot (E_0-E)^{p_2}\cdot e^{-p_3(E_0-E)},\;\;(E<E_0)$.} to describe the continuum spectrum.
The mean value of the second Gaussian is shifted by $-1\sim-2$~keV due to missing charge.
The fraction of the second Gaussian is in general less than 15\%.
Figures~\ref{fig:Am} (b) and \ref{fig:Co} (b) show the magnified energy spectra around the peak and the fit results.

Table~\ref{tab:resolution} summarizes the results of the energy resolution measurements.
Energy resolution is worse than the intrinsic noise performance of the system predominantly due to uncertainties of the gain calibration.
Gain uncertainty can also explain the fact that the energy resolution is worse for higher energy.

This result demonstrates that the energy resolution of 1~keV is an achievable goal for the SCMT, considering the planned improvements discussed below.

\section{Conclusions and future prospect}
We have fabricated prototype modules for a DSSD system which is a crucial element of the SCMT.
Intrinsic noise performance is measured to be 1.0~keV (FWHM) at 0\degC\ in the DC configuration, which is in good agreement with the analytically calculated noise value of 0.9~keV.
The energy resolution is measured to be 1.3~keV (FWHM) in the same configuration, demonstrating that the energy resolution of 1~keV is within our reach.

It is still of great importance to establish that the multiple-Compton technique can be used to measure the photon direction and polarization, as well as the energy by using a stacked DSSD setup.
This is the logical step for us to pursue next.

Moreover, we also plan to develop a full-size DSSD module based on the information obtained with the present prototype.
The size of the DSSDs will be 5 cm $\times$ 5 cm, which is the largest size possible with 4-inch wafers.
This size translates into twice the capacitance load for amplifier in the DC configuration.
Since the noise performance is dominated by the amplifier noise, improvements in the amplifier performance and reduction of the load capacitance is essential to achieve 1~keV energy-resolution.
Further improvement in the amplifier noise performance is expected by further optimizing the front-end MOSFET geometry in the 0.35~$\mu$m process\footnote{The current MOSFET geometry is optimized in the AMS 1.2~$\mu$m process.} or by employing a JFET as the front-end FET instead of a MOSFET.
We also plan to employ a thicker DSSD to reduce the capacitance to the backside.

In conclusion, we have demonstrated that energy resolution of a DSSD system is as low as 1.3~keV with the possibility of the future improvements.
Such an ultra low noise DSSD system presents great possibilities for future hard X-ray and gamma-ray telescopes, such as the SMCT.

\acknowledgments     

This work has been carried out under support of U.S. Department of Energy, contract DE-AC03-76SF00515,
Grantin-Aid by Ministry of Education, Culture, Sports, Science and Technology of Japan (12554006, 13304014),
and ``Ground-based Research Announcement for Space Utilization'' promoted by Japan Space Forum.

\bibliography{spie-2002-4851-104}   
\bibliographystyle{spiebib}   


\end{document}